## Using Artificial Intelligence to Identify State Secrets


Renato Rocha Souza (a), Flavio Codeço Coelho (a), Rohan Shah (b), Matthew Connelly (b)

(a) Escola de Matemática Aplicada, Fundação Getulio Vargas. Praia de Botafogo, 190 Rio de

Janeiro - RJ, Brasil. 22250-900

(b) Department of History, Columbia University, 1180 Amsterdam Avenue, New York, NY 10027



**Abstract:**

Whether officials can be trusted to protect national security information has become a matter of

great public controversy, reigniting a long-standing debate about the scope and nature of official

secrecy. The declassification of millions of electronic records has made it possible to analyze

these issues with greater rigor and precision. Using machine-learning methods, we examined

nearly a million State Department cables from the 1970s to identify features of records that are

more likely to be classified, such as international negotiations, military operations, and high-level

communications. Even with incomplete data, algorithms can use such features to identify 90%

of classified cables with <11% false positives. But our results also show that there are

longstanding problems in the identification of sensitive information. Error analysis reveals many

examples of both overclassification and underclassification. This indicates both the need for

research on inter-coder reliability among officials as to what constitutes classified material and

the opportunity to develop recommender systems to better manage both classification and

declassification.


**Main Text:**





More than one hundred countries have adopted laws or administrative measures giving citizens a right to information about what governments do in their name. But certain categories are typically excluded, whether because of national security or personal privacy. Distinguishing between what citizens are entitled to know and what officials are obliged to withhold is therefore a growing challenge, one that is compounded by the rapid growth in the volume of government records. In the U.S. and other countries records managers and archivists are struggling to cope, especially because electronic records like e-mail require new methods for processing and preservation (Moss 2012). This quantitative growth and qualitative transformation poses profound questions about the nature of democratic accountability in the age of "big data." As the conditions for research in historical records change, it will require rethinking how we study international relations (Allen and Connelly 2016).

The Freedom of Information Act (FOIA) request that revealed former Secretary of State Hillary Clinton's private e-mail server has put these challenges in stark relief. The FBI concluded that Clinton was "extremely careless" in overlooking sensitive information and failing to preserve public records (Comey 2016). Clinton's defenders argue that what is classified is "almost random," and that her practices were not unlike those of her predecessors (Lowell 2016). Both sides agree that it took too long to review (and redact) the 54,000 pages of e-mail she turned over to authorities, only for the FBI to re-open the inquiry ten days before the election when additional e-mail emerged. In response to another FOIA request for the e-mail of three of Clinton's top aides, which amounted to 450,000 pages, the State Department estimated that it would take seventy-five years to review all the material (LoBianco 2016). But this is just the tip of the iceberg. It is thought that the State Department generates two billion e-mail every single





year (McAllister 2010*).*

Current declassification methods are clearly inadequate to cope with the volume of potentially sensitive information that is now being generated (Public Interest Declassification Board 2012). But it is likely that many of these e-mail along with other electronic records will be lost before anyone has a chance to review them. The Office of the Inspector General found that the e-mail of many senior officials from just a few years ago may never be recovered because of lost passwords or corrupted files (State Department OIG 2016).

What can we reasonably expect from government officials charged with protecting sensitive information, and is it possible to develop systems that might help officials manage both classification and declassification more rapidly and more reliably? To date, there has been virtually no government investment in research that might answer these questions. The U.S. spent over $16 billion in 2015 to protect classified information. Just the increase in spending on secrecy over the last two years -- $4.5 billion -- is nearly ten times more than the government spends on administering the Freedom of Information Act (Department of Justice 2015), and more than ten times as much as the entire budget of the U.S. National Archives (ISOO 2015). But we have yet to see a single controlled experiment to determine to what extent officials agree on what information should be classified.

Theorists have long speculated as to the reasons why even democratic regimes invest so much more in secrecy than transparency. Excessive secrecy, or "overclassification," has long been recognized as contrary to the ostensible purpose of protecting sensitive information. It dilutes the meaning of classification and undermines respect for security procedures (Commission on





Protecting and Reducing Government Secrecy 1997). In recent years, the persistence of high levels of secrecy -- notwithstanding the Obama administration's pledge to be the most transparent in history -- has renewed interest in whether a preference for concealment is intrinsic to bureaucracy (Sagar 2013, Pozen 2010). As Max Weber famously argued:

> This superiority of the professional insider every bureaucracy seeks further to increase through the means of *keeping secret* its knowledge and intentions. Bureaucratic administration always tends to exclude the public, to hide its knowledge and action from criticism as well as it can. (Weber 1922)

In his classic history of official secrecy in Britain, David Vincent points out that Weber was really describing ideal types. Historical research reveals that not *every* bureaucracy has sought to maximize secrecy, and the amount of secrecy within the same bureaucracy can ebb and flow over time. (Vincent 1999) In 1997, during a period of relatively greater transparency in U.S. foreign policy following the end of the Cold War, the landmark report of the Moynihan Commission on Government Secrecy also drew on historical research in concluding that secrecy is best seen as a form of regulation (Commission on Protecting and Reducing Government Secrecy 1997). The problem since the very founding of the Republic is that, unlike other forms of regulation, there is no effective mechanism for determining whether the Executive is managing secrecy appropriately (Hoffman 1981; Sagar 2013).

A key challenge is therefore in determining what kinds of information really do require safekeeping, especially since the evidence from the past is largely anecdotal. Developing a richer, more robust theory of official secrecy that can both account for variation in classification





practices and inform more effective regulation requires empirical research. As Vincent acknowledges, while qualitative research in archives can reveal how officials keep secrets, it is less suited to revealing the larger patterns in what kinds of information they typically conceal. Computational analysis of declassified records can identify both patterns and anomalies, and show whether practices are consistent enough as to become predictable. This is both a good test of theory and the essential precondition for any technology that would assist in the management of sensitive information. When it comes to the Clinton e-mail, such methods can show what is normal and what might be considered negligent in how officials manage large numbers of potentially sensitive communications.

The release of the first generation of U.S. government electronic records presents new opportunities to analyze this problem using well-developed methods from natural language processing (NLP) and machine-learning. We report on the performance of algorithms developed to automatically identify sensitive information. We trained and tested these algorithms using some one million diplomatic cables from the 1970s that were originally marked as "secret," "confidential," "limited official use," or "unclassified." Cables were the main form of secure communications in this period, and since the original metadata include not only the classification level, but also the sender, recipient, date, and subject, these cables -- and research analyzing these cables -- are quite relevant to management challenges for newer types of communications, notably e-mail. They can show whether classification has historically been random or predictable, and analyzing the errors machines make in identifying sensitive information can help us understand the errors that humans make.

This article begins with a discussion of the materials and methods used in the experiment. We





then identify the features most characteristic of classified communications. We analyze how different theories of official secrecy can help explain the data, and how computational methods yield new insights. We then present the results obtained when we apply these insights to different classification tasks, depending on whether we define secrecy more or less broadly. Finally, we discuss the implications of our findings both in terms of the recent Clinton e-mail controversy and future prospects for technology that could better manage both classification and declassification.

## Materials and Methods:

Our task was to use features from the metadata and the message text to predict whether a given document would be classified or unclassified. In essence, we take a set of observations in which the category membership is known to create a training set. We then use this data to train a statistical model that scores the probability of new observations taken from a random sample to belong to the same category. Ironically, in data science this is known as a classification task.

We obtained our data from the Central Foreign Policy Files (CFPF) of the U.S. State Department, which are available from the U.S. National Archives in the form of XML files on DVDs. The CFPF was also the source of the classified diplomatic cables released by Wikileaks in 2010-11. But the best available estimate is that the 251,000 Wikileaks cables constitute only 5-6% of all the cables produced in 2005-10 (Gill and Spirling, 2015). Even that estimate requires assuming that the release constituted a random sample. But it seems unlikely that a random sample of U.S. diplomatic cables would feature more mentions of the Wrangel Islands than Russia.





There are also gaps in the declassified documents from the 1970s, and neither corpus includes the relatively few documents classified as "Top Secret." But the great advantage of using declassified historical data is that we can analyze the lacunae using both the appraisal records of archivists and the metadata that is available even for corrupted or still-classified records.

The National Archives appraisal of the electronic records of the CFPF shows how archivists and records managers endeavored to preserve all historically significant records among the 27 million that had accumulated by 2006. This corpus was both quantitatively and qualitatively different from the paper records archivists were accustomed to dealing with. They analyzed it by utilizing the 188 subject "TAGS" the original State Department drafters had assigned to each record since the system was created in 1973. This metadata field, which also included TAGS for each country, was intended to facilitate "Traffic Analysis by Geography and Subject." Archivists used it to identify and preserve all records from the 93 subject TAGS belonging to the broad categories of Political Affairs, Military Affairs, Social Affairs, Economic Affairs, Technology and Science. For the 95 other subject TAGS on things like Operations, Administration, and Consular Affairs, they reviewed a sample that totalled over seven thousand cables to decide which TAGS merited preservation even if the cables were not cross-referenced with other, more obviously significant subject TAGS.

The CFPF is therefore a curated corpus, but the effect of prioritizing the preservation of historically significant records was to reduce the relative proportion that was unclassified. Of the top forty TAGS ranked by the relative number of cables classified "secret," only one was from the Operations, Administration, or Consular Affairs categories, and that was for the





Administration of the State Department's Bureau of Intelligence and Research (AINR.)
Even these records, which archivists decided to retain, were relatively mundane -- travel reservations and the like. The largest groups of cables that archivists decided not to preserve were those related to Visas, Personnel, and Financial Management (Langbart 2007).

In training the algorithms to identify classified communications, we also had to exclude non-cable records that did not include the message text in digital form, such as paper records, many of which were delivered via diplomatic pouch ("p-reel" records). In addition, there are 410,539 "withdrawn" cables where only the metadata has been declassified. We also had to exclude cables that had no text or just "n/a" for the features we used in building the classifier (see Table 5, and see also the Supplementary Information).

The available metadata shows that about the same percentage of the withdrawn cables were originally classified as secret as compared to the cables that were released (5.2% versus 5.3%). The reason is that many records are withheld because they contain personally identifiable information and not national security information. Even fewer P-reel documents were originally classified secret (2.6%). Conversely, when we identified features in the metadata characteristic of the most secret communications, we found little difference in the rankings whether or not we included metadata from the withdrawn cables, and whether these communications were cables or p-reel records.

But there is one category of missing data that is quite distinct in terms of secrecy: 128,026 cables were meant to be preserved because they have TAGS deemed to be historically significant, but there is no message text for these cables in the State Department Automated





Data System -- only metadata. According to the U.S. National Archives, the State Department

lost some of the data during migration between different hardware and software platforms,  but

"some telegrams were intentionally deleted from the electronic repository." (National Archives

and Records Administration 2016). There are intriguing patterns in the missing data, to be

discussed below.

For purposes of the experiment, the main problem with incomplete data is that it increases the

difficulty of developing algorithms to automatically identify classified communications. But since

we still have the metadata even for these withdrawn and incomplete records, we can use them

to begin analyzing what kinds of information is typically classified. We will do that by identifying

the features in the metadata that are most useful in predicting which cables are secret, such as

who sent or received the cable, what topic it concerned, and what keywords the authors used to

categorize it. We then compute which embassies, topics, and keywords had the highest

proportion of secret cables.

In preparing the remaining 981,083 cables for the machine classification experiment, we carried

out many standard NLP operations to process the raw textual data, starting with tokenization.

Compound names of places in textual fields were aggregated, enabling them to be treated as a

single token (i. e. NEW YORK was transformed to NEWYORK). This step was especially

important in the case of the *from* and *to* fields, which represent the names of the embassies.

These fields were aggregated in a new field, *sender/recipient*, for the vectorization process. We

also eliminated all the trailing punctuation and words with length of 1. While underscores and

hyphens in the middle of words were maintained, hyphenation was otherwise eliminated from

textual fields. So too were stopwords using the NLTK (Bird 2009) English stopwords list.





Some of the available metadata indicated how these records were reviewed before release. But to test the feasibility of a system to recommend the appropriate classification under real world conditions, we only used data that would be available to a human -- or a computer -- before a cable is sent. In addition to the message text (the *body* field), we found the following features most useful in distinguishing more classified cables from the rest of the corpus: 1) *sender/recipient*, a combination of the "from" and "to" fields, which typically is an embassy or office; 2) *concepts,* standard keywords used to categorize each document; 3) *subject*, a short description; 4) *office*, the part of the State Department with responsibility for either originating or acting on a cable; 5) *TAGS*, for Traffic Analysis by Geography and Subject.

For each one of the features, and many feature combinations, we applied different forms of vectorization and tested the prediction power of the resulting vectors using several types of classifiers and ensembles. Most machine learning algorithms require fixed length vectors as inputs. For bodies of text, vectorization steps were performed using a Bag of Words (BoW) (Harris 1954) approach, using plain term counts or some form of weighting such as TfIdf (Term Frequency - Inverse Document Frequency, Spärck 1972).

All these vector transformations have parameters: the size of the vocabulary; the n-gram range taken in consideration (single tokens are unigrams; two consecutive tokens are bigrams, etc.); the choice of whether to eliminate stopwords; the minimum and maximum document frequencies; among others specific to each feature. We undertook extensive tests and hyperparameter optimization (Bergstra 2012) to determine the best parameters (Table S.4).





For this corpus TfIdf did not perform better than plain term frequencies in the BoW vectors. Despite having bigger vocabulary sizes, the fields *subject*, *concepts,* and *body* have been represented using a vocabulary that only considers the top N words ordered by term frequency across the corpus. For the *TAGS* and *Office* fields, all distinct tokens that appeared in more than five cables were taken into account. For the new field *sender/recipient,* we have discarded those that appeared in fewer than six cables, which can be indicative of misspellings.

We have used cross validation to test the performance of different models and avoid over-fitting, i.e. choosing a model that would perform poorly on data outside the sample. This entailed setting aside a random selection of cables and -- after training our classifier with the rest of the data -- testing the accuracy of its predictions for the randomly selected cables. We used the stratified k-fold method, maintaining the proportion of classes ( "unclassified," "limited official use," "confidential," "secret") in each fold. The k-fold itself uses k folds for training and test sets. We have chosen three folds for each classifier, trained with the training set and evaluated with the test set.

Rather than using just one classifier for this task, we tested twelve different classification models to evaluate their overall performance in terms of the AUC\ROC metric (Hanley 1982, Davis 2006, and see SI). Based on these tests, we built an ensemble of the seven best-performing algorithms using different weights for each: Stochastic Gradient Descent, Logistic Regression, Ridge, Bagging, Extremely Randomized Trees, AdaBoost and Multinomial Naive Bayes. This can be likened to assembling experts and giving them more or less weight depending on their comparative reliability.





Upon determining that using all features as independent vectors produced the best results, we applied this method to four different tasks: (U)nclassified vs (L)imited Official Use, (C)onfidential, and (S)ecret; U, L vs C, S; U, L, C vs S; and U vs C,S (Table 4). One could make further adjustments to the parameters, but we do not believe they would bring more than marginal gains, and the generalizability of the model would be compromised. Instead, we believe that this model could be used with similar kinds of data to achieve similar results, such as future releases from the same Central Foreign Policy Files.

## **What Makes Secret Cables Different?**

The scale and multi-dimensional nature of data from formerly classified diplomatic communications illustrates the need to use machine-learning methods to identify patterns and anomalies for otherwise hard-to-observe phenomena (Monroe et al. 2015). In this experiment, there were 40,700 possible features for each of the 918,083 records (Table S4). This much data precludes a purely qualitative analysis. Instead, it rewards an approach that allows for multiple combinations of variables and differently weighting each variable depending on the task at hand. And once we determine the features that are most useful in identifying classified records, we can look more closely at each one to indicate the kind of patterns that combine to create accurate predictions.[1]

One feature we did *not* find to be useful in predicting whether a record was more or less likely to be classified was the month and year it was dispatched. This is partly because it is difficult to

---

[1] This part of the analysis is limited to cables grouped by origin/destination, TAGS, and concepts, since other useful fields -- like Office, and the words in the message -- were not available for records that were withdrawn or where the message text is missing.





divide up the data temporally, since patterns of activity change with weekends and holidays, and some units -- like months -- are of variable length. But it's also because there was no clear trend or pattern in the proportion of classified diplomatic communications. The overall level of secrecy was little changed during the first two years of the Carter administration as compared to the time that Henry Kissinger was Secretary of State. (See Fig. 1) This is surprising, considering how the context of diplomacy in the period 1973-78 was becoming increasingly hostile to official secrecy, after the revelations of the secret bombing of Cambodia, Watergate, and the Church Committee hearings on covert CIA operations. Shortly after assuming office in 1977, Jimmy Carter famously rejected Kissinger's penchant for secrecy, promising "a foreign policy that the American people both support and, for a change, know about and understand." Echoing Woodrow Wilson, he insisted that "Our policy must shape an international system that will last longer than secret deals. We cannot make this kind of policy by manipulation. Our policy must be open; it must be candid…" (Carter 1977)

In fact, over the period 1973-78 communications regarding the same kinds of subjects continued to be classified: military matters and international negotiations. The posts that transmitted the highest percentage of secret cables include officials negotiating the Strategic Arms Limitation Treaties and the Mutual and Balanced Force Reduction agreements, those reporting from or to military staff headquarters, and those working at two embassies -- Cairo and Tel Aviv -- that were a focus of U.S. diplomacy leading up to the Camp David Accords (see Table 1).

Until recently, even the strongest proponents of transparency conceded that diplomacy must sometimes be conducted in secret, particularly when it comes to the use of force and statecraft.





Woodrow Wilson himself conducted the negotiations for the Versailles Treaty behind closed doors even after promising "open covenants of peace, openly arrived at" in his famous Fourteen Points. (Nicholson 1964) As François de Callières noted three centuries ago in a classic guide to diplomatic practice, secrecy is "the life of negotiations." (De Callières 1716) Even before the founding of the United States, the Continental Congress deemed the dispatches of the first American diplomats to be secret by default. The Constitutional Convention -- itself conducted in secret -- sought a strong executive who could act with secrecy and dispatch. (Hoffman 1981) Jeremy Bentham argued that transparency made government more efficacious, but agreed it should be "suspended" if it were calculated "To favour the projects of an enemy." (Bentham 1843) While railing against the "growing evil" of official secrecy, Giuseppe Mazzini conceded it was necessary to "Let diplomacy have its secrets, for diplomacy is but a refined mode of modern warfare..." (Mazzini 1844) Diplomats have defended secrecy by arguing that confidential negotiations are an *alternative* to warfare, and far preferable (Nicholson 1964).

The same pattern of protecting records related to international negotiations and military operations is evident in an analysis of the 862 Concepts, or keywords, State Department officials used to organize diplomatic cables (see Table 2). But so too is the tendency -- already evident in cables sent to the White House -- for the most senior officials to pursue special protection for their own communications. Ironically, the already notorious problem of overclassification made this more difficult. It was not just outside critics like Jimmy Carter who complained about the problem of excessive secrecy. Richard Nixon expressed frustration at the way even categories like "top secret" had become so overused as to lose real value or meaning. He favored creating a new designation indicative of an even higher level of exclusivity (Blanton 2003)





In 1974 the State Department created special designations for communications involving the most senior officials. Cables with the "CAT-C," "CAT-B," and "CAT-A" concepts continued to be highly classified during the Carter administration. Conversely, Concepts associated with few or no secret cables (<1%) are indicative of the kinds of subjects -- like "music" and "meats" -- that senior officials considered less important or less urgent (see Table 3).

There is also evidence that certain records were removed from the State Department Automated Data System. According to the National Archives, many of the cables where we have complete metadata but no message text are available on microfilm. Moreover, we do not know when the data were migrated, and the electronic versions of messages were lost. But it's notable that most of these cables do not date to when the State Department first set up the system, when one might expect it would have been troubleshooting ways to reliably transfer data between different hardware and software platforms. Instead, most date to 1975-76, and coincide with some of Henry Kissinger's most controversial actions at a time in which he was coming under increasing criticism for his conduct as Secretary of State.

Intriguingly, the cables missing from the database also tend to be more highly classified, and often involve the most senior officials. Electronic messages classified as "Secret" were more than three times more likely to go missing compared to Unclassified and Limited Official Use messages (22% versus 6.5%). As for the CAT-C cables, we only have the electronic message text for 38%. For the rest, there is only an error message, e.g.:





MRN: 1975JAKART014946 SEGMENT NUMBER: 000001 EXPAND ERROR ENCOUNTERED; TELEGRAM TEXT FOR THIS SEGMENT IS UNAVAILABLE

This particular record, an account of a meeting in Jakarta between Gerald Ford, Henry Kissinger, and Indonesian president Suharto, survived and was printed out in hard copy. But there are almost no State Department cables in the database from December 1-15, 1975, the first of several conspicuous gaps. When the cable was finally declassified in full in 2001, after the fall of the Suharto regime, it showed that Ford gave a green light to Suharto's stated plan to conquer East Timor using U.S. arms. Realizing this would be in violation of American law, Kissinger advised Suharto to construe it as self-defense and delay the operation until Ford had returned home. But, he said, it was particularly important that "whatever you do succeeds quickly." (Jakarta to State 1975)

Kissinger was incensed when, later that month, a U.S. foreign service officer sent a cable that apparently confirmed this violation of U.S. law. Kissinger worried about how many people had already seen it. "That will leak in three months and it will come out that Kissinger overruled his pristine bureaucrats and violated the law….Everything on paper will be used against me." (Memorandum of Conversation 1975) This second cable is one of the 119 cables sent from Jakarta that same month which are now missing from the database.

Other notable gaps in the electronic record include March 18-31 1976, when Kissinger supported the military coup in Argentina; May 25-31 1976, when he favored the Syrian invasion of Lebanon; and June 1976, when he met with the Prime Minister of South Africa in the midst of the Soweto Uprising against Apartheid. Of course, Kissinger and his staff were dealing with





many different subjects during these same periods. Even if it can be established that someone deliberately deleted the messages from the database, the data does not permit us to impute specific motives. But it is notable that the gaps end with the end of Kissinger's term as secretary of State (See Fig. 2).

Since Bentham and John Stuart Mill, philosophers have long postulated that secrecy serves to cover up the abuse of power, and that liberty requires transparency. As Louis Brandeis famously observed, "sunlight is...the best of disinfectants." (Brandeis 1914) In analyzing the relationship between these different features, we find intriguing patterns that provide some empirical support for these beliefs. For instance, the first two years of the Carter administration witnessed rapid growth in the number of cables with the SHUM TAGS, designating those related to Human Rights. This was consistent with the Carter administration's stated policy of promoting democratic values. There was also tremendous growth in the number of human rights organizations. Some, like Freedom House, also reported on the situation in each country, giving them a 'Freedom in the World Score' from 1 (best, e.g. Norway) to worst (7, e.g. North Korea). The index is calculated by grading the country across 10 political rights indicators, such as the electoral process. There are also 15 civil liberties indicators, such as freedom of expression, that are broadly derived from the Universal Declaration of Human Rights.

When we compare the percentage of cables diplomats wrote about human rights when reporting about different countries in 1977-78 (i.e. by counting the cables in which country TAGS co-reference SHUM TAGS, and comparing it to all cables with these country TAGS) we find only a very weak relationship with the score these countries received from Freedom House (an r of 0.2108108261.) In other words, whether U.S. diplomats were more or less concerned





with the human rights situation in a country had little to do with the human rights situation per se, at least as it was assessed by Freedom House. In fact, U.S. diplomats largely ignored the human rights situation in many allied countries with repressive governments. This is contrary to what outside critics like Jean Kirkpatrick argued at the time. They claimed that Carter had a blanket policy of criticizing human rights violations, which made little difference with Cold War adversaries and only served to delegitimate allied governments (Kirkpatrick, 1979).

In fact, there is a striking regional variation in how much U.S. diplomats wrote about human rights -- even more striking when we limit the analysis to countries that imported U.S. arms in this period. Middle Eastern and Latin American countries both tended to have repressive governments, with an average Freedom House score of 4.7 and 4.5 respectively. But there was far more reporting about the human rights situation in Latin America compared to the Middle East. Of cables concerning Argentina, Bolivia, Brazil, Chile, Colombia, Ecuador, Paraguay, Peru, Uruguay, and Venezuela, 5.96% reference Human Rights. For Egypt, Iran, Israel, Jordan, Kuwait, North Yemen, Saudi Arabia, Turkey, and the UAE, it's <1% (.85%.) (See Fig. 3)

These findings are consistent with earlier research by Qian and Yanagizaw showing a divergence between the annual State Department human rights reports and the assessments of Amnesty International. They found that the State Department showed favoritism towards its Cold War allies, which they identified according to how consistently other countries voted with the U.S. at the U.N. General Assembly. But they also control for regional diversity in their assessment (Qian and Yanagizaw 2009). We find that, in fact, the specific situation of allied countries in the Cold War is what is most strongly associated with whether the State Department showed any interest in human rights, and not just whether it criticized human rights violations in





published reports. When countries were well within the U.S. sphere of influence, and there was little risk of Soviet intervention, such as Latin America in 1973-78, foreign service officers frequently reported on detention, torture, and civil liberties. But they showed little interest in such issues in Middle Eastern countries, where the U.S. and the USSR were engaged in intense competition for influence and the risk of international conflict was high.

Not coincidentally, countries of the Middle East were also far more likely to have a high percentage of secret cables. Almost fifteen percent (14.82%) of all cables with Middle Eastern country TAGS were classified as secret in 1973-78. For Latin American Countries, it was two percent (2.26%)

**Results:**

The great advantage of using a machine-learning approach to identify what specific records are more likely to be classified is that it can leverage many subtle interrelationships in multidimensional data. In this case, we used it to identify cables originally classified as Secret; Confidential; Limited Official Use; and Unclassified, which indicate the degree of sensitivity accorded to these communications by the officials who drafted them. We grouped them in various ways to measure the performance of different classifiers depending on how broadly or how narrowly one defines a state secret. Of all the features, the relative frequency of different words in the *body* was the most useful in identifying sensitive information. High recall *or* precision (but not both) is achievable with some of the other features. For instance, it is possible to identify 95% of the cables that are classified as Secret, Confidential, or Limited Official Use just by using the *sender/recipient* data. But fully a third of the identified cables would be false





positives. Other features produce better overall performance (see Fig. 4). But the best results came from combining all feature as independent vectors.

Table 4 shows several performance measures when we categorize sensitive information more or less broadly. This can be assessed in terms of accuracy (Walther 2005) (how often is the classifier correct?), recall (also known as true positive rate: how many of the classified cables does it identify?), precision (or false positive rate: how many of the identified cables were actually classified?), and the average f1 score (the harmonic mean of precision and recall) for the two classes. We also present the AUC or ROC Score (Hanley 1982) - the area under the curve when plotting the true positive rate (y-axis) against the false positive rate (x-axis) as you vary the threshold for assigning observations to a given class. All of these metrics are derived from the true/false positive and negative counts.

There is a notable improvement in performance when we exclude data that does not clearly belong to either category. The "Limited Official Use" designation has long been a cause for complaint. White House Executive Orders from the period classify as "Secret" information that "could reasonably be expected to cause serious damage to the national security," and include specific examples. "Confidential" was the classification for information that could cause "damage" but not "serious damage" (Federation of American Scientists 1972). Limited Official Use, on the other hand, had no such definition. When the George W. Bush administration attempted such a definition, the Government's leading FOIA expert admitted it was "so broad that it is almost harder to think of information that does not fall within its scope than information that does" (Metcalfe 2009). This is consistent with our inability to distinguish these cables despite using many different features and multiple classifiers. It is not clear whether any NLP or





machine learning methods would accurately distinguish this category.

But the results also show there is some logic in official secrecy, at least to the extent that State Department cables from the 1970s are indicative. Even when we include Limited Official Use cables, just a few kinds of metadata and the relative frequency of words in the message text are enough to identify 90% of classified cables with relatively few (<11%) false positives. Whether a diplomatic communication should be classified may therefore be relatively predictable -- more so than, say, predicting heart disease by data-mining medical records (Chaurasia 2014).

If we were able to use all the diplomatic cables from the period, including both those that were lost and those that are still classified, we would likely achieve higher accuracy. But even with all of the data, these methods will sometimes miss crucial contextual elements, such as whether the subject of a seemingly innocuous travel reservation or visa application was on a sensitive mission (See Fig. 5). Other errors might instead reflect the intrinsic subjectivity of at least some classification decisions, especially considering that officials often receive inconsistent or inadequate guidance.

When we began to examine the false positives and false negatives -- i.e. the cables we predicted would be classified, but were unclassified, and vice versa -- we found many instances of human error. For unclassified cables which the algorithm identified as having the highest probability of being classified, this included hundreds of examples where the cables were miscategorized as unclassified in the metadata, such as a report on Japanese government sensitivity about U.S. inspection of its nuclear facilities. The message text itself clearly showed it was originally confidential (Amembassy to Secstate 1977). Other errors included cables that





were originally secret when received at the State Department but were resent to another post as unclassified, such as  a report on what Lebanese Christian leaders said about ceasefire negotiations with the PLO (Secstate to Cinceur 1977). And there were many examples of unclassified cables that, according to experts with security clearances we consulted, would have been highly sensitive at the time. This included, for instance, what a confidential informant told U.S. diplomats in Cyprus about the kidnapping of the President's son  (Fig. 2) (Secstate to Usdel 1977).

**Discussion:**

There is an upper limit to what any supervised learning algorithm can achieve in machine classification: it cannot be more accurate than the intrinsic consistency of the data allows. To the extent humans misclassify their communications -- or simply disagree about how they should be classified -- so too will algorithms.

In the debate over the Clinton email, many appear to assume that the post-hoc review created what data scientists would call "ground truth" in identifying classified communications. But one could argue that Clinton and her aides were no less expert in recognizing what communications deserved to be classified and contained on a secure system. It might instead be seen as a natural experiment in inter-coder reliability in identifying state secrets. If so, this would be the government's first such experiment in 75 years of creating official secrets, or at least the first in which it has published the results.

But this experiment is flawed and incomplete, especially if it is expected to answer the question





of how careless Clinton and her aides were in handling email that were deemed to be classified. These "false negatives" constitute 6.9% of the total reviewed. Even if we accepted this as ground truth, we cannot begin to estimate either precision or recall -- and hence accuracy -- without reviewing the communications Clinton and her aides correctly classified as sensitive and protected on a secure system (true positives), and those which they *overclassified* (false positives). Moreover, one would need to compare the error rate with the error rate for the rest of the State Department. Certainly, the idea that government officials make no errors in identifying sensitive national security information is not supported by the data.

The historical data also show with new precision the extent to which officials have preserved, or failed to preserve, the official record. Whereas press coverage of the Clinton controversy has focused on very recent practices of using private e-mail, the historical record shows that diplomats have for centuries used both official and unofficial channels to control access to their communications. What has changed in recent years is that, with the use of electronic records, we can now use computational methods to analyze different kinds of communications. When records are "lost" it becomes more difficult, but these acts can also leave statistically conspicuous gaps in the public record that merit further scrutiny. Without this kind of analysis, a democracy loses the capacity to hold individuals and public institutions to account, a great asset against less democratic types of regimes that offsets their ability to act with even greater "secrecy and despatch." When instead officials do not preserve the public record a democracy is doubly disadvantaged, losing both the capacity to identify and correct errors and incompetence, while still remaining relatively open and prone to leaking compared to autocratic states (Colaresi 2012).





Based on what we have learned from machine classification, it should be feasible to develop systems in which classified and unclassified communication streams would continually generate data that would be predictive of the appropriate classification level for new communications. Such a system would automatically default to the predicted classification, requiring manual override if the sender wished to classify at a higher or lower level (such as communications about a new classified program, or a subject that is no longer sensitive).

The same type of system could be used for accelerating the release of electronic records, harnessing data from previous declassification decisions to prioritize records for close scrutiny that are most likely to have sensitive national security or personal information. These systems could therefore "nudge" officials to classify or declassify communications appropriately, and also reveal which officials tend to diverge from the norm when dealing with similar types of subjects, language, etc. Such a system could therefore be self-correcting, and become increasingly accurate as sources of error were continually identified and corrected for.

What is most needed now is government support for applied research on technology to improve management of sensitive information (Public Interest Declassification Board 2012). This should include rigorous, double-blind testing of how consistent humans are in both classification and declassification. Without determining the expected error rate among humans, we will never know the baseline against which to test computer-assisted approaches -- or evaluate claims that this or that official was negligent in failing to recognize and protect state secrets.

**Acknowledgements**

This research was made possible through grants from the John D. and Catherine T. MacArthur Foundation, the Columbia Global Policy Initiative, and the Fundação Getulio Vargas. The authors also thank Eric Gade and Thomas Nyberg for expert assistance, and David Blei, Michael Gill, Richard Immerman, Robert Jervis, Daniel Krasner, Aaron Plasek, Owen Rambow, and Arthur Spirling for helpful discussions.






**Fig. 1:**

Proportion of Cables Classified as Secret 1974-1978

*The total includes all cables, including those that are still withheld, dating from the first year in which we have relatively complete data.*

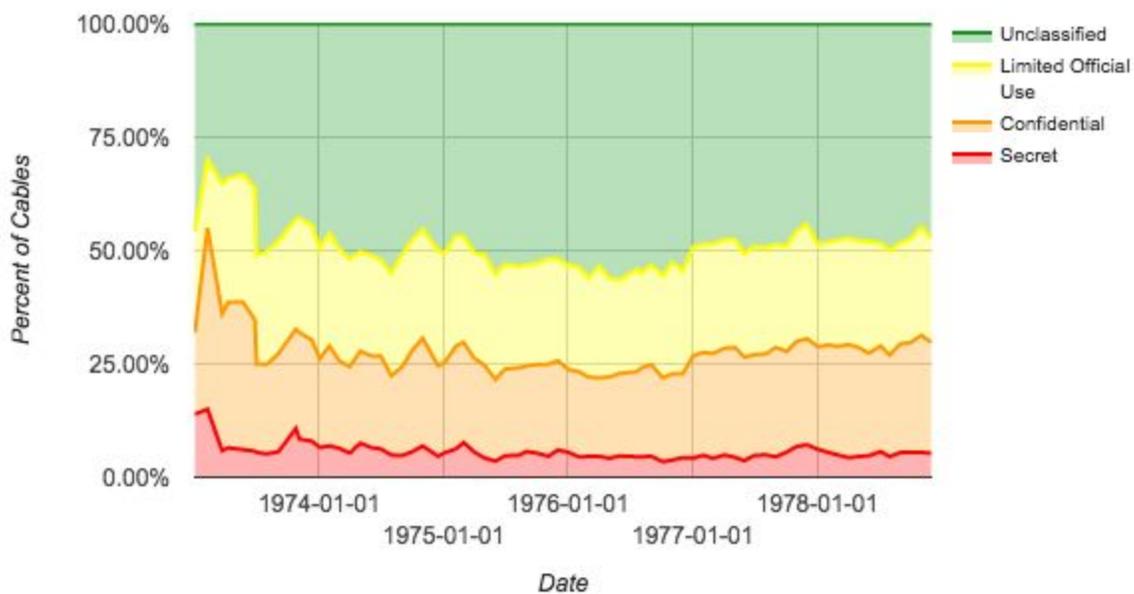

**Fig. 2:**

Number of Missing Cables, 1973-1978

*The number of cables by day with error messages in the place of message text.*





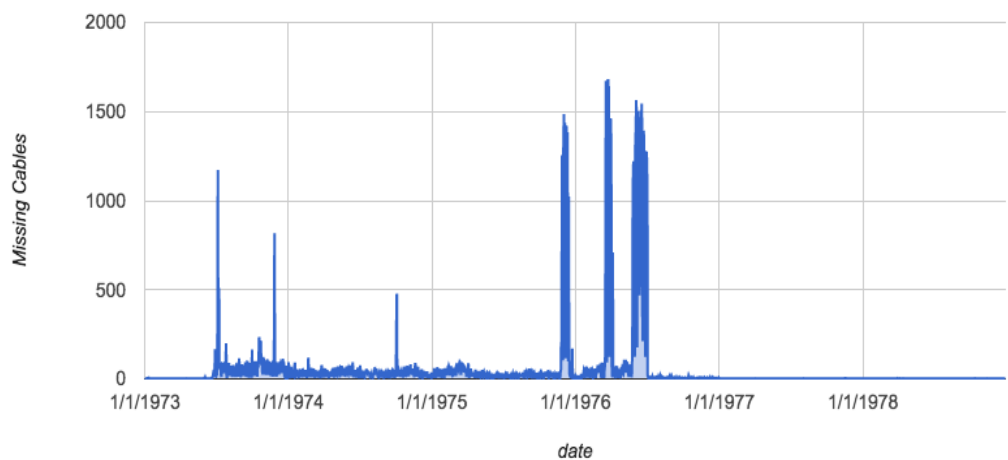

**Fig. 3:**

Percentage of Cables by Country Related to Human Rights and Percent Classified as Secret

*Countries with a severe human rights situation according to Freedom House are color-coded orange or red. The size of the bubble is proportionate to the number of cables with different country TAGS. The Y axis is a log scale showing which countries have a higher proportion of cables related to human rights. The X axis shows which countries have a higher proportion of cables classified as secret.*





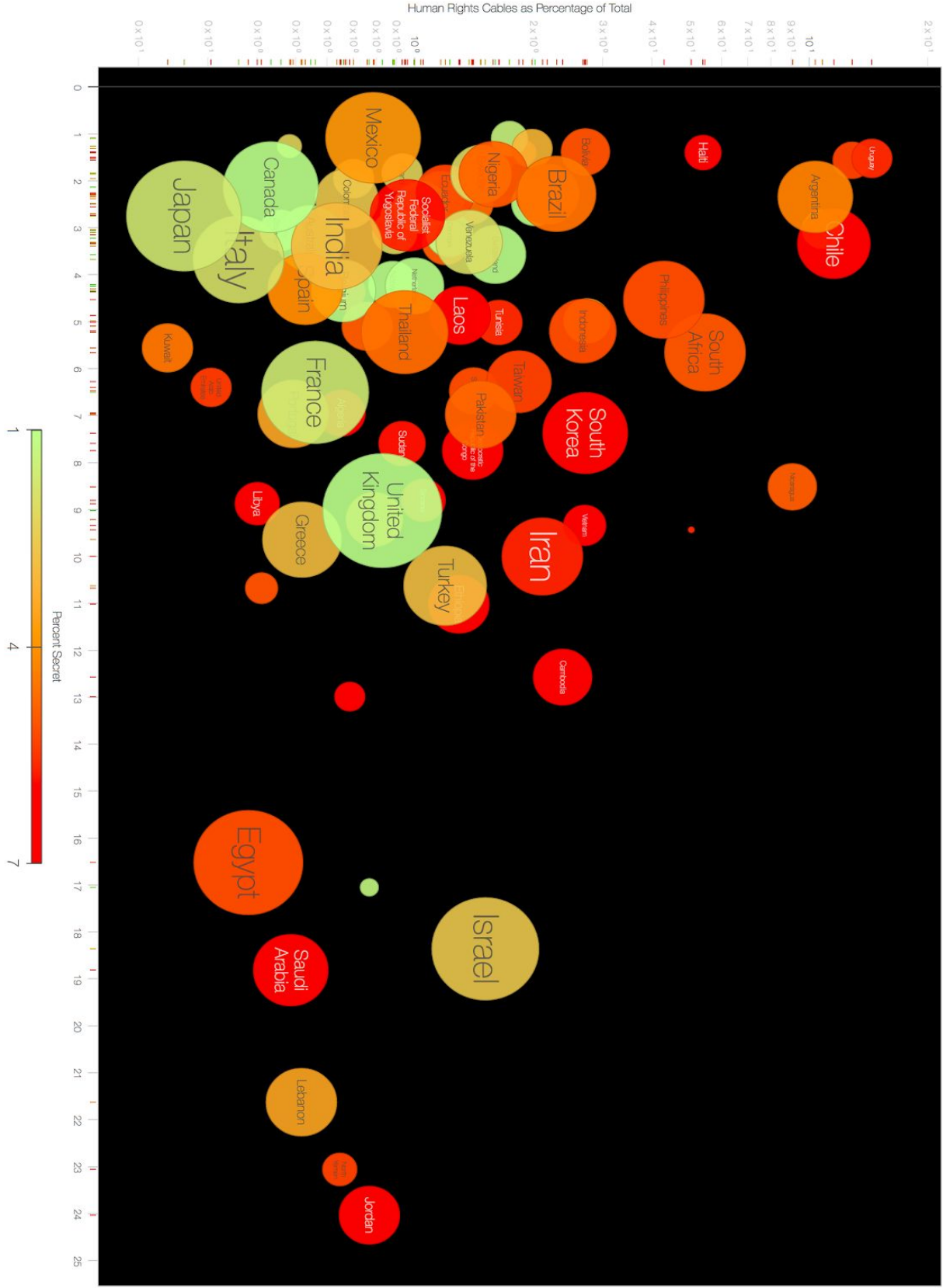

Human Rights Cables as Percentage of Total





**<u>Fig. 4:</u>**

Performance of Classifiers Using Different Features

*To determine which features are most useful, we tested the performance of classifiers for each one individually.*

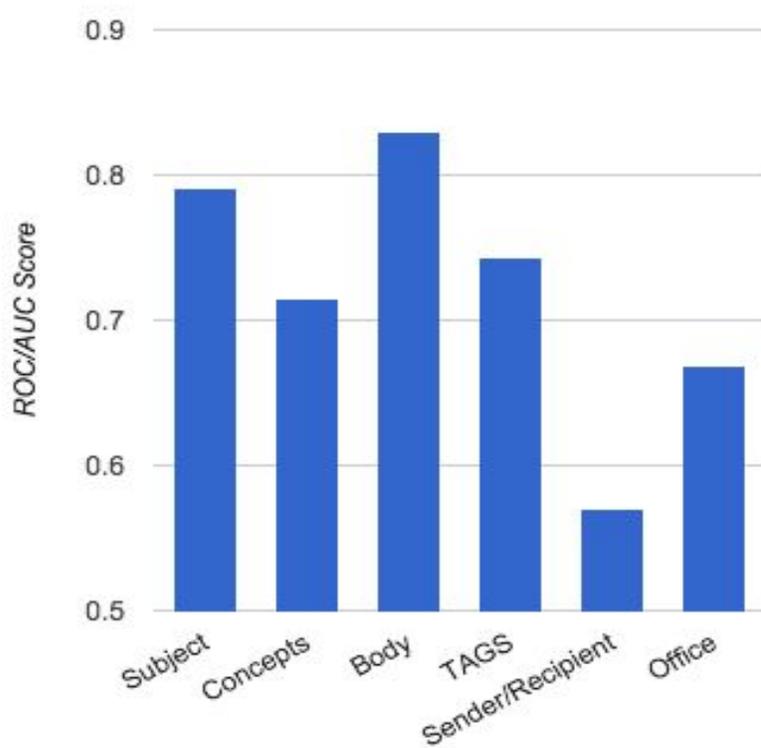

**<u>Fig. 5:</u>**

How We Measure Performance in Identifying State Secrets

*Classifiers can be optimized for either recall or precision. When the end-user has low tolerance*





*for risk, they will seek to maximize recall, i.e. the percentage of all cables with sensitive*

*information they identify as classified, even if the result is lower precision, i.e. the percentage of*

*all cables identified as classified that actually have sensitive information. But the false positives*

*and false negatives include examples of human error in classification.*

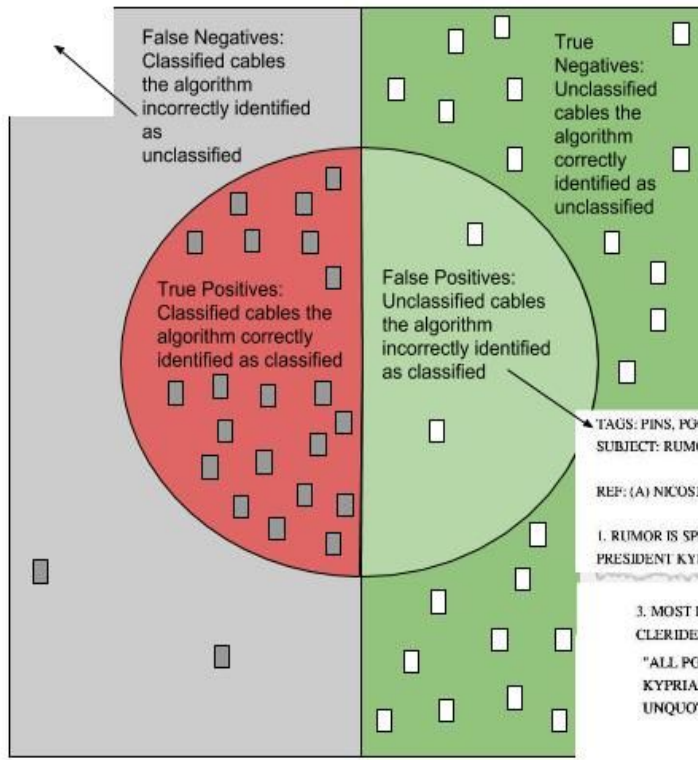





**Table 1:**

Origin and Destination of Cables Most Likely To be Classified as Secret

*After filtering out the from/to pairs that did not send at least one hundred secret cables from*

*1973-78, these are the ones that had the highest percentage of secret cables overall.*

| From | To | Secret Cables | Total Cables | Percent Secret |
|------|-----|--------------|--------------|----------------|
| NATO | STATE, SECDEF, INFO: MBFR VIENNA, BONN, LONDON, USNMR SHAPE, USCINCEUR | 426 | 436 | 97.71% |
| CAIRO | STATE SS MULTIPLE | 104 | 109 | 95.41% |
| SALT TALKS | STATE | 671 | 723 | 92.81% |
| GENEVA USSALTTWO | STATE | 434 | 486 | 89.30% |
| TEL AVIV | STATE SS MULTIPLE | 145 | 163 | 88.96% |
| STATE | SECRETARY WHITE HOUSE | 121 | 146 | 82.88% |
| MBFR VIENNA | STATE DOD | 775 | 1075 | 72.09% |
| STATE | WHITE HOUSE | 1464 | 2035 | 71.94% |
| STATE | JCS MULTIPLE | 179 | 250 | 71.60% |
| STATE | USCINCEUR | 131 | 258 | 50.78% |

**Table 2:**

Concepts (or Keywords) With the Highest Proportion of Secret Cables

*After filtering out the concepts that did not appear in at least 1,000 cables from 1973-78, these*

*are the ones that had the highest percentage of secret cables overall.*





| Concept | Secret Cables | Total Cables | Percent Secret |
|---|---:|---:|---:|
| CAT-C | 6,211 | 7,156 | 87.00% |
| CAT-B | 9,377 | 11,821 | 79.00% |
| MUTUAL FORCE REDUCTIONS | 1,800 | 2,466 | 73.00% |
| SALT (ARMS CONTROL) | 1,950 | 2,990 | 65.00% |
| CAT-A | 3,234 | 5,158 | 63.00% |
| FORCE & TROOP LEVELS | 1,228 | 2,036 | 60.00% |
| INTERIM GOVERNMENT | 909 | 1,745 | 52.00% |
| MISSILES | 1,069 | 2,336 | 46.00% |
| PEACE | 845 | 1,838 | 46.00% |
| PEACE PLANS | 1,738 | 4,025 | 43.00% |

**Table 3:**

Concepts With the Lowest Proportion of Secret Cables

*These are the more common "concepts" (>1,000 total cables) assigned to cables in which officials deemed <1% merited classification as "secret."*

| Concept | Secret Cables | Total Cables |
|---|---:|---:|
| CIVIL AVIATION | 75 | 16,715 |
| SCIENTIFIC VISITS | 41 | 11,758 |
| TEXTILES | 7 | 7,157 |
| MUSIC | 26 | 6,439 |
| SEMINARS | 25 | 5,756 |
| EXCEPTIONS LIST | 13 | 5,312 |
| SCIENTIFIC MEETINGS | 11 | 5,081 |
| LABOR UNIONS | 24 | 5,009 |
| ENVIRONMENT | 11 | 4,763 |
| MEATS | 5 | 4,213 |





**Table 4:**

Performance of Classifiers Using All Six Features for Different Tasks.

| Classifica-tion Task | ROC/ AUC Score | Accuracy Score | Precision Score for Un-classified | Precision Score for Classified | Recall for Un-classified | Recall for Classified | Average f1-score |
|---|---|---|---|---|---|---|---|
| (U vs L,C,S) | 0.859 | 0.87 | 0.834 | 0.891 | 0.81 | 0.9 | 0.896 |
| (U,L vs C,S) | 0.85 | 0.87 | 0.884 | 0.843 | 0.92 | 0.78 | 0.809 |
| (U,L,C vs S) | 0.806 | 0.966 | 0.974 | 0.802 | 0.99 | 0.62 | 0.697 |
| (U vs C, S) | 0.928 | 0.929 | 0.926 | 0.931 | 0.94 | 0.92 | 0.926 |

**Table 5:**

Counts and Classification Types of Cables with Information on Cables Considered Blank

| Situation | Total in Database | Unclassified | Limited Official Use | Confidential | Secret |
|---|---|---|---|---|---|
| **declassified cables** | **1,758,279** | **876,797** | **411,973** | **375,690** | **93,635** |
| Error messages for *body* | 119,744 | 53,935 | 21,744 | 25,233 | 18,832 |
| blank *body* | 8282 | 2,726 | 1,645 | 1,924 | 1,987 |
| blank or n/a *concepts* | 634,967 | 445,300 | 114,507 | 65,502 | 9,658 |
| blank or n/a *subject* | 26,109 | 16,490 | 5,820 | 2,914 | 885 |
| blank or n/a *from* | 17 | 7 | 6 | 3 | 1 |
| blank or n/a *to* | 9,740 | 6,027 | 1,572 | 1,698 | 443 |
| **Used for classifier** | **981,083** | **368,043** | **280,251** | **270,477** | **62,312** |





**Supporting Information**

**Data acquisition and preparation**

The Central Foreign Policy Files available from the US. National Archives include cables as well as so-called p-reel records, i.e. communications sent physically via diplomatic pouch. Both are either available in full or have been "withdrawn" because the record either has sensitive national security information or personal information. Both p-reel records and withdrawn cables have limited metadata, but neither has message text. We limited our analysis to cables that have been fully released.

The metadata includes the original classification (*origclass*), i.e. Secret, Confidential, Limited Official Use and Unclassified. Cables with null, degenerated or misspelled names of classes in *origclass* were left out of the analysis.

**Table S1:**
Cable Counts and Classification According to Cable Type

| *Cable_type* field | Total in Database | Unclassified | Limited Official Use | Confidential | Secret | Null or degenerated |
|---|---|---|---|---|---|---|
| Full cables | 1,758,279 | 876,797 | 411,973 | 375,690 | 93,635 | 184 |
| P-reel | 505,030 | 419,052 | 33,887 | 28,695 | 14,716 | 0 |
| Withdrawn cables | 410,539 | 298,932 | 25,988 | 25,816 | 21,474 | 38,329 |
| P-reel withdrawn | 8,920 | 3,538 | 570 | 2,277 | 2,304 | 231 |





We first analyzed the cables to decide which fields could be useful as input features. The feature engineering involved analysis of the textual quality; discovering most common values for each class; and defining the subset valid for the tests. The fields that were added *a posteriori* to the creation of the cable were not used as they could convey information related to the classification. The main feature of interest is the *body* - the full text of the cable - although we have also dealt with other fields (Table S2).

**Table S2:**

Explanation of fields used in Classifier

| Used Fields | Description |
| --- | --- |
| *origclass* | The original classification level of the cable |
| *body* | Full text of the cable |
| *subject* | Topic dealt with in the document. |
| *concepts* | Keywords attributed to the document |
| *TAGS* | Traffic Analysis by Geography and Subject |
| *from* | Which person/what office sent the document. |
| *to* | Which person/what office received the document. |
| *office* | Which State Department office or bureau was responsible for the document. |
| *date* | Document creation date |

Not all the cables that present *cable_type* field equal to *full* have useful content in the *body*. Some of them present errors in the digitization process and have, as their bodies, just small error messages, as illustrated in Table S3.





**Table S3:**

Overview of Error Types in Full Cable Dataset and Count of Classification

| Type of errors in digitization | Total in Database | Unclassified | Limited Official Use | Confidential | Secret |
|---|---|---|---|---|---|
| "Error Reading Text Index" | 46,876 | 13,931 | 6,241 | 11,592 | 15,112 |
| "Expand Error Encountered" | 72,850 | 39,988 | 15,499 | 13,640 | 3,723 |
| "Encryption Error" | 42 | 22 | 10 | 10 | 0 |
| Total (errors are not exclusive) | 119,744 | 53,935 | 21,744 | 25,233 | 18,832 |

The tests aimed at predicting the level of sensitivity in a binary fashion; therefore, we have tested binary aggregations (called here scenarios) of the four classes grouping them according to broad or narrow definitions of secrecy: *(S)ecret, (C)onfidential, (L)imited Official Use* and *(U)nclassified*. Additional tests using four classes as targets were also performed, but the nature of the cables' subjects and classification led to low recall and precision measurements.

Once the features were selected, the database was queried, and the retrieved data were joined in a Python Pandas (McKinney 2010) Dataframe structure, we processed the raw textual data by eliminating hyphenation, tokenizing (separating words and discarding punctuation), and removing common, non-informative words such as "and", "the", etc. Tests were made using stemmed forms of words, but it didn't enhance the performance and the stemming was discarded. The field *date* was transformed in a boolean field *weekday* - indicating whether the date fell in a weekend or not; and another field *year+month*, used to test hypothesis on the





temporal nature of classification (i.e. a larger proportion might be classified during weekends, or periods of crisis). This did not prove useful for the whole span, although it could be promising for analysis of shorter periods of time.

## Feature vectors

For each one of the features, and some feature combinations, we have applied a few forms of vectorization and tested the prediction power of the resulting vectors. Like Blei et al. (Blei 2003), we define the following terms: A word or token is the basic unit of discrete data, defined to be an item from a vocabulary indexed by {1,...,N}. Words are represented using vectors that have a single component equal to one and all other components equal to zero. A document vector is a sequence of N word counts denoted by $w = (w_1, w_2, ..., w_N)$, where $w_n$ is the nth word in the sequence and N is the size of the vocabulary. A corpus is a collection of M documents denoted by $D = \{[w_{11}, w_{21}, ..., w_{N1}], [w_{12}, w_{22}, ..., w_{N2}], ..., [w_{1M}, w_{2M}, ..., w_{NM}]\}$.

With Bag of Words (BoW), documents are represented by vectors of dimension N, where N is the size of the vocabulary or the subset N of the most frequent words of the vocabulary; each column represents a word and the value is the frequency of that word in the document. TfIdf (Term Frequency - Inverse Document Frequency) also represents documents as vectors of dimension N, where N is the vocabulary size or the subset N of the most high-valued words in the vocabulary, based on the TfIdf metric: instead of word counts in the columns, it utilizes a method for emphasizing words that occur frequently in a given document (Tf), whilst de-emphasising words that occur frequently in many documents or in the whole corpus of documents (Idf). The count of the words in the vector is substituted by some weighting factor





(Salton 1988); in our case it is $ \mathrm{tfidf}(t,d,D) = \mathrm{tf}(t,d) + \mathrm{tf}(t,d) \times \mathrm{idf}(t, D) $ where t is the term, d is the document, and D is the corpus of documents. In our particular implementation, we have discarded terms that appeared in less than 2 documents, eliminating some of the misspellings and hapaxes.

We undertook extensive tests and hyperparameter optimization (Bergstra 2012) to determine the best parameters, and summarize the results in Table S.4.

**Table S.4:**

Overview of Parameters of Analysis against Feature Type

| Feature | Number of Tokens | Vocabulary size (N) | Max vector size (N') | Best n-gram range | Best vectorization |
|---------|------------------|---------------------|----------------------|-------------------|--------------------|
| *subject* | 6,894,992 | 180,480 | 8,000 | (1,1) | Term frequencies |
| *concepts* | 4,929,265 | 13,192 | 650 | (1,2) | Term frequencies |
| *body* | 259,276.062 | 1,929,902 | 15,000 | (1,1) | Term frequencies |
| *TAGS* | 3,272,125 | 939 | 844 | (1,1) | Term frequencies |
| *Embassy (from/to)* | 2,234,457 | 4,874 | 1,036 | (1,1) | Term frequencies |
| *office* | 1,937,707 | 261 | 170 | (1,1) | Term frequencies |
| All Text | 278,544,608 | 1,968,680 | 15,000 | (1,1) | Term frequencies |

**Classifiers and Ensembles**





We performed extensive tests with different classifiers: Linear Models (Logistic Regression, Passive Aggressive, Stochastic Gradient Descent, Ridge Regression; Perceptron); Support Vector Machines (Linear SVC); K-Nearest Neighbors; Bayesian Approaches (Bernoulli Naive Bayes, Multinomial Naive Bayes); ensembles of classifiers (Random Forests, Extremely Randomized Trees) and combined techniques such as bagging and boosting (Bagging Classifier, Gradient Boosting, AdaBoost and weighted voting approaches). See supplementary references for more on these different approaches.

In order to choose the best classifiers for the task, a prediction comparison was made in advance, using all textual features combined and the scenario [U (class 0) vs L,C,S (class 1)]. Each one of them had its parameters optimized for the task with grid search. The results for BoW/TfIdf vectorization scores are shown in Table S5.

**Table S5:**
Performance of All Classifiers Across Various Measures

| Classifier | ROC/AUC Score | Accuracy Score | Precision (class 0/1) | Recall (class 0/1) | F1-score (class 0/1) |
|---|---|---|---|---|---|
| Stochastic Gradient Descent | 0.8462/ 0.8338 | 0.8576/ 0.8512 | (0.82/0.88)/ (0.83/0.86) | (0.80/0.89)/ (0.76/0.90) | (0.81/0.89)/ (0.79/0.88) |
| Logistic Regression | 0.8457/ 0.8434 | 0.8569/ 0.8574 | (0.81/0.88)/ (0.82/0.88) | (0.80/0.89)/ (0.79/0.90) | (0.81/0.89)/ (0.81/0.89) |
| Linear SVM* | 0.8454/ 0.8452 | 0.8563/ 0.8583 | (0.81/0.88)/ (0.82/0.88) | (0.80/0.89)/ (0.79/0.90) | (0.81/0.89)/ (0.81/0.89) |
| Ridge | 0.8261/ 0.8373 | 0.8448/ 0.8546 | (0.82/0.86)/ (0.83/0.87) | (0.75/0.90)/ (0.77/0.91) | (0.78/0.88)/ (0.80/0.89) |
| Bagging (w/ Dec. Tree) | 0.8048/ 0.8049 | 0.8172/ 0.8173 | (0.76/0.85)/ (0.76/0.85) | (0.76/0.85)/ (0.76/0.85) | (0.76/0.85)/ (0.76/0.85) |
| Extremely Randomized Trees | 0.8036/ 0.7938 | 0.8365/ 0.83 | (0.86/0.83)/ (0.86/0.82) | (0.67/0.94)/ (0.65/0.94) | (0.76/0.88)/ (0.74/0.87) |
| AdaBoost (w/ Random F.) | 0.8031/ 0.8072 | 0.8190/ 0.8222 | (0.77/0.85)/ (0.77/0.85) | (0.74/0.87)/ (0.75/0.87) | (0.75/0.86)/ (0.76/0.86) |
| Random Forest | 0.7964/ 0.7994 | 0.8310/ 0.8316 | (0.86/0.82)/ (0.85/0.82) | (0.66/0.94)/ (0.67/0.93) | (0.74/0.87)/ (0.75/0.87) |
| Perceptron* | 0.7856/ 0.7963 | 0.8138/ 0.8112 | (0.80/0.82)/ (0.75/0.84) | (0.67/0.90)/ (0.74/0.86) | (0.73/0.86)/ (0.75/0.85) |
| Passive Aggressive* | 0.7745/ 0.8226 | 0.8095/ 0.837 | (0.82/0.81)/ (0.79/0.86) | (0.63/0.91)/ (0.76/0.88) | (0.71/0.86)/ (0.78/0.87) |
| Multinomial Naive Bayes | 0.7735/ 0.7828 | 0.7614/ 0.7992 | (0.64/0.87)/ (0.74/0.83) | (0.82/0.73)/ (0.72/0.85) | (0.72/0.79)/ (0.73/0.84) |
| Bernoulli Naive Bayes | 0.6885/ 0.6885 | 0.6538/ 0.6538 | (0.52/0.84)/ (0.52/0.84) | (0.83/0.55)/ (0.83/0.55) | (0.64/0.66)/ (0.64/0.66) |

Some classifiers did better with BoW vectorization and others with TfIdf. Based on these tests, we chose an ensemble with the seven best estimators (Stochastic Gradient Descent, Logistic Regression, Ridge, Bagging, Extremely Randomized Trees, AdaBoost and Multinomial Naive Bayes) and using weights 2,2,1,1,1,1,1 for our main experiment.

**Evaluation of Results**

Table S6 presents the performance of the classifier for the three combinations of classes using each feature as input as well as two feature combinations. The fields *body*, *subject*, *concepts*, *tags*, *embassy* and *office* were added together in a new field/feature *all_text*, which was tested as an alternative to combining all the other features by concatenating those vectors. All vectors were produced using plain count BoW. All measures are calculated as a mean of the three folds using stratified k-fold.





**Table S6:**
Performance of Classifier in Different Capacities by Feature Type

| Feature | Class Combination | ROC/AUC Score | Accuracy Score | Precision (class 0/1) | Recall (class 0/1) | Average f1-score |
|---|---|---|---|---|---|---|
| Subject | (U vs L,C,S) | 0.79 | 0.82 | 0.81/0.82 | 0.68/0.91 | 0.74/0.86 |
| | (U,L vs C,S) | 0.80 | 0.83 | 0.85/0.77 | 0.89/0.72 | 0.87/0.74 |
| | (U,L,C vs S) | 0.70 | 0.96 | 0.99/0.80 | 0.99/0.40 | 0.98/0.53 |
| Concepts | (U vs L,C,S) | 0.72 | 0.75 | 0.69/0.77 | 0.59/0.84 | 0.63/0.81 |
| | (U,L vs C,S) | 0.74 | 0.78 | 0.80/0.74 | 0.89/0.58 | 0.84/0.65 |
| | (U,L,C vs S) | 0.68 | 0.91 | 0.96/0.75 | 0.99/0.36 | 0.97/0.48 |
| Body | (U vs L,C,S) | 0.83 | 0.84 | 0.79/0.87 | 0.78/0.88 | 0.79/0.87 |
| | (U,L vs C,S) | 0.81 | 0.84 | 0.85/0.82 | 0.92/0.70 | 0.88/0.75 |
| | (U,L,C vs S) | 0.68 | 0.95 | 0.96/0.76 | 0.99/0.36 | 0.98/0.49 |
| TAGS | (U vs L,C,S) | 0.74 | 0.78 | 0.75/0.79 | 0.61/0.88 | 0.67/0.83 |
| | (U,L vs C,S) | 0.75 | 0.79 | 0.82/0.72 | 0.87/0.63 | 0.84/0.67 |
| | (U,L,C vs S) | 0.62 | 0.95 | 0.95/0.73 | 0.99/0.25 | 0.97/0.38 |
| Embassies (From/To) | (U vs L,C,S) | 0.57 | 0.67 | 0.71/0.66 | 0.19/0.95 | 0.30/0.78 |
| | (U,L vs C,S) | 0.59 | 0.69 | 0.70/0.65 | 0.93/0.24 | 0.80/0.35 |
| | (U,L,C vs S) | 0.57 | 0.94 | 0.94/0.72 | 1.00/0.14 | 0.97/0.24 |
| Office | (U vs L,C,S) | 0.67 | 0.73 | 0.76/0.72 | 0.42/0.92 | 0.54/0.81 |
| | (U,L vs C,S) | 0.62 | 0.73 | 0.71/0.83 | 0.97/0.27 | 0.82/0.41 |
| | (U,L,C vs S) | 0.62 | 0.95 | 0.95/0.79 | 1.00/0.25 | 0.97/0.38 |
| All_Text | (U vs L,C,S) | 0.85 | 0.86 | 0.82/0.88 | 0.81/0.89 | 0.81/0.89 |
| | (U,L vs C,S) | 0.84 | 0.87 | 0.88/0.84 | 0.92/0.76 | 0.90/0.80 |
| | (U,L,C vs S) | 0.78 | 0.92 | 0.97/0.78 | 0.99/0.57 | 0.98/0.66 |
| All Features | (U vs L,C,S) | 0.86 | 0.87 | 0.83/0.89 | 0.81/0.90 | 0.82/0.90 |





| (independent vectors) | (U,L vs C,S) | 0.85 | 0.87 | 0.88/0.84 | 0.92/0.78 | 0.90/0.81 |
| | (U,L,C vs S) | 0.81 | 0.97 | 0.97/0.80 | 0.99/0.61 | 0.98/0.69 |
| | (U vs C, S) | 0.93 | 0.93 | 0.93/0.93 | 0.94/0.92 | 0.93/0.93 |

Upon determining that using all feature as independent vectors produced the best results, we applied this method to a corpus in which all the cables are clearly identified as classified or unclassified, i.e. with no Limited Official Use (L) cables in either the training or test set. The ROC Score of ~0.93 -- the best result of all the tasks -- indicates how these ambiguous cables limit the performance of classifiers for identifying state secrets. The misclassified cables -- both false positives and false negatives -- were also presented to two domain experts with the classification removed. They were more likely to agree with the algorithm in judging cables to have information that would have been sensitive in the era they were created.

**Figure S1:**

AUC scores for all features

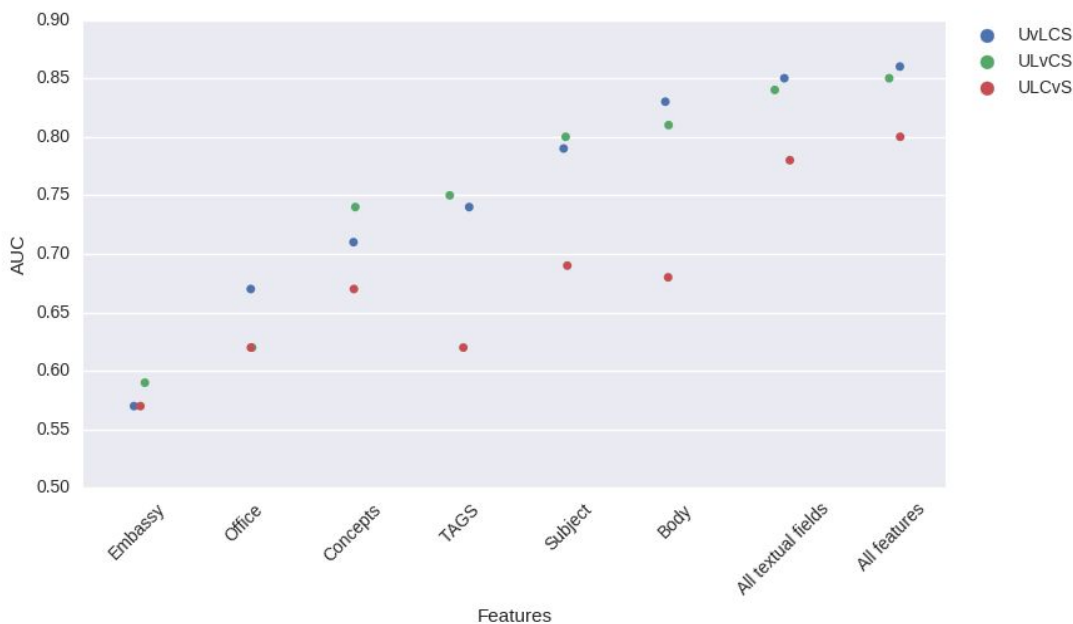





**Figure S2:**
Precision vs. recall for each feature and the combination of features

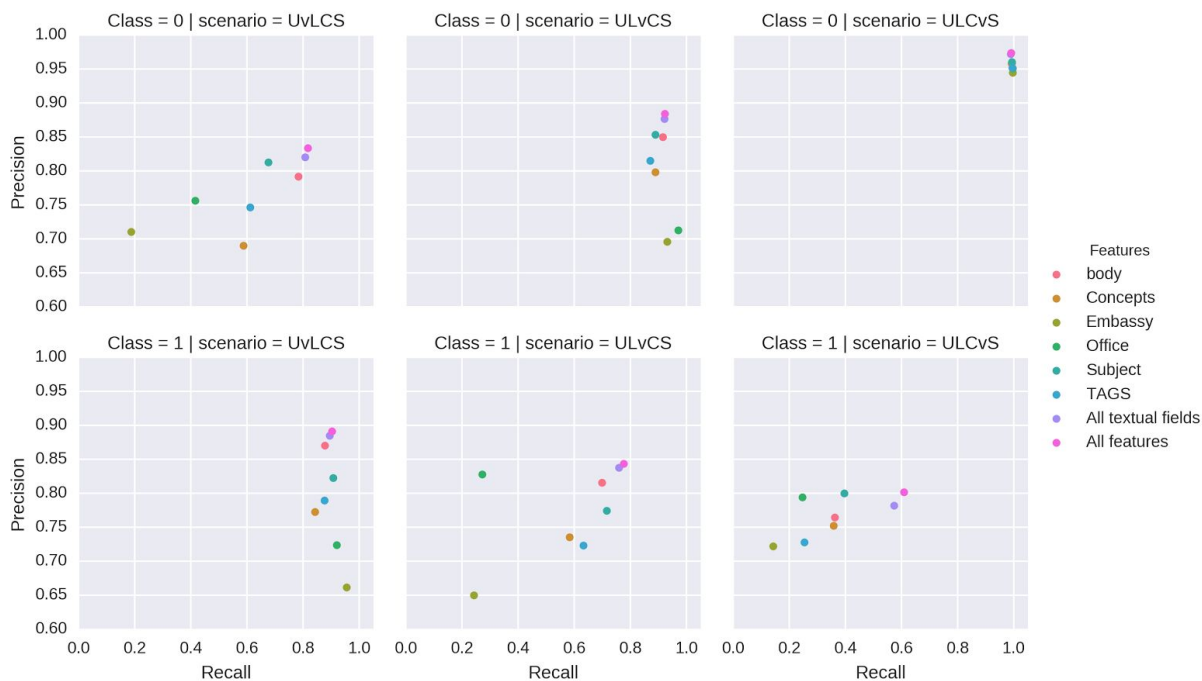

Using all features yielded the best results in all scenarios, closely followed by the summation of textual fields, then body and subject. The other features did not show good performance taken alone.

## Discussion

Results over 80% in classification tasks are generally regarded as good (e.g. the Netflix challenge, in which the best result was ~85%)(Netflix 2009). But whether such results are adequate depends on the task at hand. If our priority is to protect sensitive information, and we wish to minimize the number of false negatives, i.e., cables that are predicted to be unclassified





but are in fact classified, the classification thresholds could be altered at the cost of increasing the number of false positives.

Our current developments point to future enhancements of the classifier. We are testing an implementation of an Artificial Neural Network, exploring new features, and developing methods to correct errors in the metadata (original classes). These methods may further improve the results.

## Supplementary References